\documentclass[prd,preprint,amsmath,nofootinbib]{revtex4}
\usepackage{graphics}

\begin{document}

\title{On  a  class of  stable,  traversable  Lorentzian wormholes  in
  classical   general  relativity}   \author{C.  Armend\'ariz-Pic\'on}
\email{armen@oddjob.uchicago.edu}
\affiliation{Enrico Fermi Institute,
  Department of Astronomy and Astrophysics,
  \\ University of Chicago.}

\begin{abstract}
  It is  known that  Lorentzian wormholes must  be threaded  by matter
  that  violates  the null  energy  condition.  We  phenomenologically
  characterize such  exotic matter by  a general class  of microscopic
  scalar field Lagrangians and formulate the necessary conditions that
  the existence of Lorentzian wormholes imposes on them.  Under rather
  general  assumptions,  these  conditions  turn out  to  be  strongly
  restrictive.  The most simple  Lagrangian that satisfies all of them
  describes a minimally coupled  massless scalar field with a reversed
  sign  kinetic  term.   Exact,  non-singular,  spherically  symmetric
  solutions  of Einstein's equations  sourced by  such a  field indeed
  describe  traversable  wormhole  geometries.   These  wormholes  are
  characterized by two parameters: their mass and charge.  Among them,
  the  zero  mass  ones   are  particularly  simple,  allowing  us  to
  analytically  prove  their   stability  under  arbitrary  space-time
  dependent perturbations.   We extend our arguments  to non-zero mass
  solutions and conclude that at least a non-zero measure set of these
  solutions is stable.
\end{abstract}

\maketitle

\section{Introduction and Summary}

The study of Lorentzian  wormholes \cite{Visser} in general relativity
has suffered from the absence of conventional microscopic descriptions
of the matter that holds open  the throat of the wormhole.  Indeed, it
is known on  one hand that such matter has to  violate the null energy
condition, while  on the  other hand, most  known matter forms  do not
\cite{MorrisThorne}.  Thus,  many wormholes have  been investigated in
alternative  theories  of  gravity,  such as  Brans-Dicke  theory,  or
quantum effects  have been invoked.  However,  in principle, classical
general  relativity  admits  stable  wormhole solutions  supported  by
simple matter  forms.  In this paper  we introduce a  general class of
matter Lagrangians  and study the  properties they have to  satisfy in
order  to  allow  the  existence  of wormholes.   These  matter  forms
necessarily violate some of the standard energy conditions, and hence,
their study also offers the possibility to address the physical nature
of  these  conditions  and  their  relation to  issues,  such  as  the
stability of vacuum.
  
A  phenomenological way of  microscopically characterizing  an unknown
form  of matter  is to  describe  it by  a scalar  field.  Whereas  an
ordinary scalar  field always satisfies  the null energy  condition, a
scalar  field  with non-standard  kinetic  terms  (different from  the
conventional  squared   gradient)  can  violate   any  desired  energy
condition \cite{ArMu}.  In  this paper we assume that  the matter that
threads  a  wormhole  consists  of  a  scalar  field  $\varphi$  whose
Lagrangian  $p$ is an  a-priori undetermined  function of  the squared
gradient  $(\nabla\varphi)^2$.  It  turns  out that  the existence  of
wormhole  solutions in general  relativity essentially  determines the
form of this  Lagrangian.  It has to describe  a massless scalar field
whose  kinetic term has  the opposite  sign as  conventionally assumed
($p=-\frac{1}{2}(\nabla\varphi)^2$              instead             of
$p=\frac{1}{2}(\nabla\varphi)^2$).   Although we  must agree  that the
issue is  yet unsettled, we  have not discovered any  inconsistency in
this  choice and  we  are not  aware  of any  physical principle  that
forbids such  a field.   In particular, we  have analysed  in Appendix
\ref{sec:appendix}  the stability of  Minkowski vacuum  against second
order perturbations in  such a theory, and our  analysis has not shown
any substantial difference to the  Minkowski vacuum in the presence of
a  conventional  massless  scalar  field.   In a  certain  sense,  the
opposite sign is preferable to the conventional one.  It is known that
all  spherically  symmetric solutions  of  Einstein's field  equations
coupled  to   a  massless   scalar  field  \cite{Fisher}   have  naked
singularities    (admittedly,    these    solutions    are    unstable
\cite{JetzerScialom}).   However,  if  the  massless scalar  field  is
coupled to gravity  with the opposite sign, most  of the solutions are
regular  everywhere,  and  describe  in  fact  wormholes  \cite{Ellis,
  Bronnikov}.   Even   cosmological  solutions  are   better-behaved.  
Instead  of running  or  originating at  a  big-bang singularity,  the
universe  bounces  at  a  finite  value  of the  scalar  factor  in  a
transition form contraction to expansion.  Obviously, the non-singular
behavior of these solutions is linked to the violation of the standard
energy conditions by such a field.

Ironically, in their quest  of non-singular particle-like solutions in
general  relativity, Einstein  and Rosen  \cite{EinsteinRosen} already
pointed out  in 1935 that  by reversing the  sign of the  free Maxwell
Lagrangian, one  can obtain  solutions free from  singularities, which
may be  in fact interpreted  as charged particles. Scalar  fields with
the opposite  sign of  their kinetic terms  have been  also previously
considered in  the literature.  As  mentioned, they have  been already
shown  to allow  wormhole solutions  \cite{Ellis,  Bronnikov, Kodama},
they appear in certain models of inflation \cite{ArMu}, they have been
proposed  as dark  energy  candidates \cite{Caldwell},  and they  also
appear in certain unconventional  supergravity theories which admit de
Sitter solutions  \cite{Hull}.  In this  work we conclude on  one hand
that  the  existence of  wormhole  solutions  forces  such a  negative
kinetic  term, but  on  the other  hand,  we argue  that although  the
physical significance and relevance of  such a field is unclear, these
wormhole  solutions are  perfectly  sensible.  For  instance, one  may
suspect  that due  to  the  unconventional form  of  the scalar  field
Lagrangian, those  wormhole solutions are unstable.   We have analysed
the stability of a particular  class of solutions, those of zero mass,
against arbitrary  linear perturbations of  the metric and  the scalar
field.  These perturbations can be decomposed into spherical harmonics
of definite ``angular  momentum'' $L$, and the stability  of each mode
$L$ can  be studied  separately.  We find  that all modes  are stable,
proving  this way  that the  zero  mass solutions  are stable  against
arbitrary linear perturbations.

The paper is organized as follows.  In Section \ref{sec:geometries} we
review  the  basic  properties  of  wormholes and  discuss  how  their
existence constrains  the forms of  matter that may support  them.  In
Section \ref{sec:k-field} this matter  is characterized by an a-priori
undetermined scalar field  Lagrangian.  The requirements of asymptotic
flatness  and  existence of  a  wormhole  throat  essentially fix  the
previously undetermined Lagrangian.  As  mentioned, it has to describe
a minimally  coupled scalar field with  a sign reversed  kinetic term. 
Section  \ref{sec:solutions} discusses  static,  spherically symmetric
solutions  of  Einstein's  field  equations  sourced  by  an  ordinary
massless scalar field.  All of them contain naked singularities at the
origin.   By assuming  the  field  to be  purely  imaginary, which  is
equivalent to assume that the kinetic term of the scalar field has the
opposite  sign,  we  rediscover  the  regular  wormhole  solutions  of
\cite{Ellis}.   They are  parametrized by  a  mass and  a charge,  and
Section  \ref{sec:stability} addresses  the stability  of a  subset of
these wormholes,  namely, those  of zero mass.   It is found  that all
solutions in  this subset are stable,  and that this  stability can be
also  extended to  sufficiently small  values of  the  mass.  Finally,
Appendix \ref{sec:appendix}  illustrates the extent  an unconventional
sign in the massless scalar  field Lagrangian affects the stability of
Minkowski space,  and concludes  that in any  case even  the Minkowski
vacuum in the presence of a  conventional scalar field is in a certain
sense gravitationally unstable.

\section{Wormhole geometries} \label{sec:geometries}
\begin{figure}
  \begin{center}
  \includegraphics{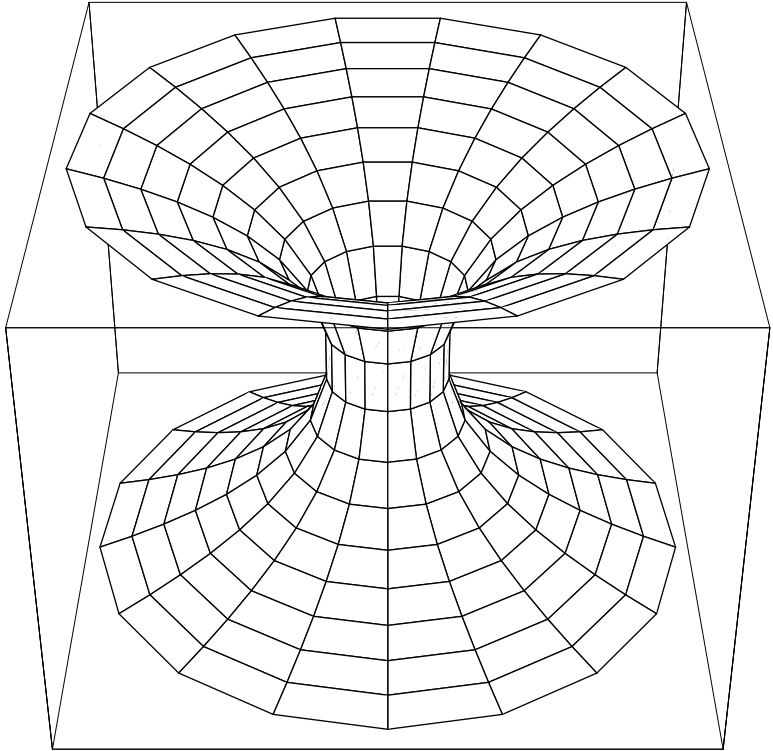}
  \caption{A spatial section $t=const$ of the wormhole
    (\ref{eq:metric}) (one dimension  suppressed).  The coordinate $l$
    labels  the  proper  distance   to  the  throat  along  ``radial''
    directions.   Surfaces of  constant  $l$ have  area $4\pi  r^2(l)$
    (shown  here as  circles of  circumference $2\pi  r(l)$).   As one
    approaches  $l=\pm\infty$,  the   location  of  the  two  distinct
    asymptotic infinities, the circumference of the circles approaches
    $2\pi |l|$, the geometry becomes flat.\label{fig:wormhole}}
\end{center}
\end{figure}

Roughly speaking, a (Lorentzian) wormhole is a spacetime whose spatial
sections  contain   two  asymptotically  flat  regions   joined  by  a
``throat''  (see  Figure  \ref{fig:wormhole}).  The  first  recognized
example of a wormhole, the Einstein-Rosen bridge \cite{EinsteinRosen},
is part of the maximally extended Schwarzschild solution of Einstein's
vacuum  field equations.   In the  former,  the presence  of an  event
horizon   prevents   observers   from   traveling  between   the   two
asymptotically flat regions.  By definition, traversable wormholes are
wormhole geometries that  do not contain horizons, in  such a way that
an  observer  may  travel in  both  ways  through  the throat  of  the
wormhole.   For   simplicity,  we  shall   consider  only  spherically
symmetric, static  wormholes in  this paper.  Any  static, spherically
symmetric metric can be cast in the form
\begin{equation}\label{eq:metric}
  ds^2=e^{2\nu(l)}dt^2-dl^2-r^2(l)\,d\Omega^2,
\end{equation}
where  $d\Omega^2=d\theta^2+\sin^2  \theta\,   d\phi^2$  is  the  line
element  on a  unit 2-sphere.   The function  $\nu(l)$  determines the
frequency  of  freely  propagating   photons  as  measured  by  static
observers, and it is hence called the redshift function.  The variable
$l$ is a proper distance  coordinate and surfaces of constant $l$ have
areas equal  to $4\pi\, r^2(l)$.   In order to describe  a traversable
wormhole, the metric (\ref{eq:metric}) has to satisfy the following
conditions:
\begin{itemize}
\item Existence of two asymptotically flat regions
  
  The  spatial  sections of  (\ref{eq:metric})  contain two  different
  asymptotically flat infinities,  if as $l\to\pm\infty$, the function
  $r^2$  approaches  $l^2$.  These  two  different  regions of  space,
  $l\to\infty$ and  $l\to -\infty$, are  connected by a  ``throat'' if
  the  function  $r(l)$  is   regular  everywhere.   Without  loss  of
  generality, we can choose the wormhole throat, the $l=const$ surface
  of minimal area, to be at $l=0$.  Then, at the throat,
\begin{equation}\label{eq:flaring-r}
r'(0)=0\quad \mathrm{and}\quad r''(0)>0,
\end{equation}
where a prime ($'$) means a derivative with respect to $l$. A spatial
geometry that satisfies these conditions is shown in Figure 
\ref{fig:wormhole}.

\item Absence of horizons
  
  So   that  no  horizons   prevent  the   passage  between   the  two
  asymptotically  flat  regions,  the  function $e^{2\nu}$  should  be
  non-zero.   At the  same time,  observers should  be able  to travel
  between two  arbitrary points  in space in  a finite  time, implying
  that $\nu(l)$ should be finite everywhere.

\end{itemize}

It turns  out, that the mere  existence of a  wormhole throat severely
constrains  the  properties  of  the  matter  that  threads  it.   Let
$T_{\alpha\beta}$  be  the stress-energy  tensor  of  the matter  that
supports the wormhole and lets denote its different components by
\begin{equation}\label{eq:components}
\rho\equiv T^t{}_t,\quad
\tau\equiv T^l{}_l,\quad
p\equiv -T^\theta{}_\theta=-T^\phi{}_\phi.
\end{equation}
Hence,  $\rho$, $\tau$  and $p$  are  the energy  density, the  radial
tension and  the ``angular'' pressure  measured by a static  observer. 
We  work  in  units where  $4\pi  G=1$  and  our metric  signature  is
$(+,-,-,-)$. The  only non-trivial  Einstein equations for  the metric
(\ref{eq:metric}) are then
\begin{subequations}\label{eq:einstein}
  \begin{eqnarray}
  \label{eq:rho-field}
  \rho&=&-\frac{r''}{r}+\frac{1-r'{}^2}{2 r^2} \\
  \label{eq:tau-field}
  \tau&=&-\frac{\nu' r'}{r}+\frac{1-r'{}^2}{2 r^2} \\
  \label{eq:p-field}
     p&=&\frac{1}{2}\left(\nu''+\nu'{}^2+\frac{\nu' r'+r''}{r}\right),
   \end{eqnarray}
\end{subequations}
which  in turn  automatically  imply  what can  be  considered as  the
equation of motion of matter
\begin{equation}\label{eq:matter}
\tau'+(\nu'+2\frac{r'}{r})\,(\tau+p)=0.
\end{equation}
If     one     subtracts     equation    (\ref{eq:tau-field})     from
(\ref{eq:rho-field}) and  evaluates the result at  the throat ($l=0$),
one gets using the finiteness of $\nu$ and (\ref{eq:flaring-r})
\begin{equation}\label{eq:null}
  (\rho-\tau)\big|_\mathrm{throat}=-\frac{r''(0)}{r(0)}<0.
\end{equation}
Recall that $\tau$ is a tension, i.e.  a negative pressure.  Hence, as
it  is well-known  \cite{MorrisThorne},  the matter  that threads  the
throat of  the wormhole violates the  null\footnote{$\rho+p\geq 0$ and
  $\rho-\tau\geq  0$}, the  weak\footnote{$\rho\geq0$,  $\rho+p\geq0 $
  and $\rho-\tau\geq 0$} and the strong\footnote{$\rho-\tau+2p\geq 0$,
  $\rho+p\geq0$    and    $\rho-\tau\geq    0$}   energy    conditions
\cite{HawkingEllis}.  We call such matter ``exotic''.

\section{Wormholes supported by a k-field}\label{sec:k-field}
One can draw two different  conclusions from the fact that only exotic
matter may  be able  to support  a wormhole. If  one assumes  that all
matter forms  satisfy the above mentioned energy  conditions, then our
previous result  excludes wormholes  from general relativity.   On the
other hand, if one insists upon the existence of wormholes, it follows
that the matter  that threads them has to be  ``exotic''. Let us point
out that,  to our knowledge, there  is no reason why  all matter forms
should satisfy the null, weak  and strong energy conditions.  In fact,
recent  indications  that   the  universe  is  currently  accelerating
\cite{Supernova} directly  imply that  the strong energy  condition is
violated  in nature, by  what seems  to be  acting as  a non-vanishing
positive cosmological constant.  The weak energy condition is violated
by a negative cosmological constant, and the study of spaces with such
a  negative cosmological constant  has recently  attracted significant
attention  (see  references to  \cite{Maldacena}).   Finally a  scalar
field with non-linear kinetic  terms, a k-field, can basically violate
any desired energy condition \cite{ArMu}.  One of the purposes of this
paper is  to show that,  at least in  principle, a simple  k-field may
indeed support wormholes in general relativity. It turns out, that the
Lagrangian  we  are  going  to  consider respects  the  arguably  only
physically  motivated  energy  condition,  the  requirement  that  the
energy-momentum  current $T^\beta{}_\alpha  u^\alpha$  measured by  an
observer    with    arbitrary     four    velocity    $u^\alpha$    be
non-spacelike\footnote{This  implies $|\rho|\geq |p|$  and $|\rho|\geq
  |\tau|$.}.

A k-field is a scalar field $\varphi$ minimally coupled to Einstein
gravity. By definition, its action is given by 
\begin{equation}\label{eq:action} 
S[g_{\alpha\beta},\varphi, \psi_m]
=\int d^4x \sqrt{-g} \left[-\frac{R}{4}+p(X)\right]+
S_m[g_{\alpha\beta},\psi_m].
\end{equation}
Here, the  k-field Lagrangian $p(X)$  is an arbitrary function  of the
squared scalar field gradient,
\begin{equation}
  X\equiv \frac{1}{2}g^{\alpha\beta}\nabla_\alpha\varphi 
  \nabla_\beta\varphi,
\end{equation}
which for  a static, spherically symmetric  configuration is negative,
$X=-\frac{1}{2}(\varphi')^2$.   For  simplicity  we  assume  that  the
k-field Lagrangian  $p$ only depends  on the squared gradient  $X$ and
not  on  the  k-field  $\varphi$  itself.  Therefore,  the  theory  is
symmetric   under   constant   shifts   of   the   field   $\varphi\to
\varphi+\epsilon$  and the  corresponding current  conservation yields
the equation of motion of the k-field,
\begin{equation}\label{eq:k-field-motion}
  \nabla_\alpha\left(p_{,X} \nabla^\alpha \varphi\right)=0,  
\end{equation} 
where  $,X$ denotes  a derivative  with  respect to  $X$.  The  action
(\ref{eq:action})   can   be   regarded   as  the   simplest   generic
phenomenological  way of  providing a  microscopic description  of the
unknown matter that supports a wormhole. We assume that the field only
couples  to Einstein  gravity,  and not  to  additional matter  fields
$\psi_m$.

The  energy-momentum tensor  of  the k-field  is  given by  functional
differentiation of the k-field action
\begin{equation}\label{eq:EMT}
T_{\alpha\beta}\equiv\frac{2}{\sqrt{-g}}
  \frac{\delta S}{\delta g^{\alpha\beta}}=
  \frac{dp}{dX}\nabla_\alpha\varphi\nabla_\beta\varphi-p\,g_{\alpha\beta}.
\end{equation}
If the scalar field gradient is space-like (as is the case in a static
field configuration) it can be cast in the form $T_{\alpha\beta}=
(\rho-\tau)\,n_\alpha n_\beta -p\, g_{\alpha\beta}$, where $n_\alpha$
is a space-like unit vector pointing in the direction of the field
gradient,
\begin{equation}\label{eq:tau}
  \tau=2 X \frac{dp}{dX}-p
\end{equation}
is the radial tension, $\rho=-p$ is  the energy density and $p$ is the
pressure as defined in  (\ref{eq:components}).  Notice that, unlike in
a  situation where the  field gradient  is time-like  \cite{ArMu}, the
energy-momentum tensor  does \emph{not} have perfect  fluid form.  The
k-field  equation of motion  (\ref{eq:k-field-motion}) can  be derived
both  by  functional  differentiation  of  the k-field  action  or  by
requiring the energy momentum tensor to be covariantly conserved.  For
a  static,  spherically   symmetric  field  configuration,  the  field
equation    is   given    by   the    ``energy    conservation''   law
(\ref{eq:matter}).

The different properties of  a wormhole severely restrict the possible
Lagrangians   $p(X)$    which   may   support    it.    According   to
(\ref{eq:null}), the  null energy condition  has to be  violated at
the  throat of  the wormhole.   Using  the properties  of the  k-field
energy momentum tensor, one finds  that $\tau+p$ should be positive at
$l=0$ and hence  there should exist an $X_T$, the value  of $X$ at the
throat, such that
\begin{equation}\label{eq:flaring}
\frac{dp}{dX}\Big|_{X_T}<0.
\end{equation}

On the other  hand, the requirement of asymptotic  flatness also yields
important information about the form  of the pressure $p$ at infinity. 
The gravitational field of  any isolated source in asymptotically flat
spacetime can be expanded  in multipole moments.  If the
source  is  static and  spherically  symmetric  only monopole  moments
contribute  to  the  gravitational  field,  and there  exists  then  a
coordinate system  where the metric admits an  asymptotic expansion of
the form \cite{Thorne}
\begin{equation}\label{eq:multipoles}
  ds^2=\left(1+\frac{\phi_1}{r}+\frac{\phi_2}{r^2}+\cdots\right)
  dt^2-\left(1-\frac{\psi_1}{r}-\frac{\psi_2}{r^2}-\cdots\right)
  \left[dr^2+r^2 d\Omega^2\right].
\end{equation}
Of  course,   the  k-field  cannot   be  considered  as   a  localized
(finite-size) source.  We assume  that the k-field decays sufficiently
fast   at    infinity,   in   such   a   way    that   the   expansion
(\ref{eq:multipoles})  still  applies.    Plugging  the  above  metric
expansion coefficients  into Einstein's field equations  (we shall not
write them down here), one can derive analogous expansions for $p$ and
$\tau$,
\begin{eqnarray*}
  p&=&-\frac{3\psi_1^2+8\psi_2}{8\, r^4}+\mathcal{O}(r^{-5}) \\
  \tau&=&\frac{\phi_1-\psi_1}{2\,r^3}
  -\frac{4\phi_1^2-8\phi_2-2\phi_1\psi_1+5\psi_1^2+8\psi_2}{8\,r^4}
  +\mathcal{O}\left(r^{-5}\right).
\end{eqnarray*} 
Substituting  these expansions into  the field  equation of  motion in
these coordinates, one finds  $\phi_1=\psi_1$ (as appropriate for a mass
term) and $2\phi_2=\psi_1^2$. Therefore,
\begin{equation}
  p=-\frac{3\psi_1^2+8\psi_2}{8\, r^4}+\mathcal{O}(r^{-5})=
\tau+\mathcal{O}(r^{-5}).
\end{equation}
At  infinity  both pressure  and  radial  tension  vanish and  in  the
asymptotic limit $r\to\infty$ the ratio  of the pressure to the radial
tensor tends to 1.  We conclude that there should exist an $X_\infty$,
the asymptotic value of $X$ as $l\to\infty$, where
\begin{equation}\label{eq:vanishing-pressures}
  p(X_\infty)=0,\quad \tau(X_\infty)=2 X_\infty \cdot p_{,X}(X_\infty)=0
\end{equation}
and such that, around $X_\infty$,
\begin{equation}\label{eq:ultrahard}
  \frac{p(X_\infty+\Delta X)}{\tau(X_\infty+\Delta X)}\approx
  \frac{p_{,X}(X_\infty)}{\tau_{,X}(X_\infty)}=1.
\end{equation}
We  will  argue below  that  $\tau_{,X}$  should  be non-zero  between
$X_\infty$ and  $X_T$. Then, in  order for (\ref{eq:ultrahard})  to be
satisfied,  $p_{,X}(X_\infty)$ should  be  also different  from zero.  
Therefore,   it  follows   from   (\ref{eq:vanishing-pressures})  that
$X_\infty=0$, and thus, around the origin
\begin{equation}\label{eq:linear}
p=\tau\approx c\cdot X,
\end{equation}
where $c$ is an arbitrary  non-zero coefficient which can be chosen to
be $\pm 1$ by a field  redefinition.  Hence, up to a sign, the k-field
behaves at  infinity as  a massless scalar  field. This  is consistent
with the pressure decaying as $1/r^4$ in the limit $r\to\infty$, which
corresponds  to a  $1/r$ decay  of the  field $\varphi$  at  infinity. 
Finally, as pointed out above, a well defined field equation of motion
requires
\begin{equation}\label{eq:regular}
  \tau_{,X}\neq 0 \quad \mathrm{for}\quad  X_T\leq X\leq X_\infty=0.
\end{equation}
The reason  is that the  points where $\tau_{,X}=0$ are  barriers over
which  the  field  equation  of  motion  (\ref{eq:matter})  cannot  be
integrated,  so that  a continuous  field  configuration interpolating
between  $X_T$  and $X_\infty$  requires  $\tau_{,X}$  to be  non-zero
within that interval.
 
\begin{figure}
  \begin{center}
  \includegraphics{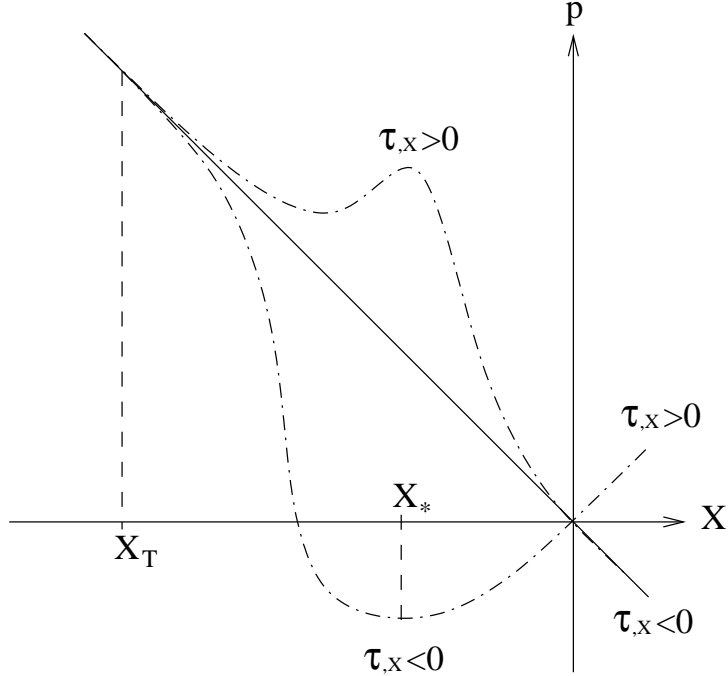}
  \caption{Possible forms of the Lagrangian $p$. It
    must have a negative slope at  $X_T$, it must be linear around the
    origin but  because of  (\ref{eq:regular}), it cannot  have local
    maxima or minima.  Thus, $p$  has to be a monotonically decreasing
    function    of    $X$   with    a    negative    slope   at    the
    origin.\label{fig:p-of-x} }
  \end{center}
\end{figure}

Conditions       (\ref{eq:flaring}),       (\ref{eq:linear})       and
(\ref{eq:regular})  essentially determine the  form of  the Lagrangian
$p$ (as  illustrated in  Figure \ref{fig:p-of-x}).  Around  the origin
$p$ must  be linear and, in  addition, there must exist  a point $X_T$
where the slope of $p$  is negative.  Suppose that $p\approx X$ around
the origin.   Then there must exist  a local minimum  of $X_*$ between
$X_T$    and    $X=0$    where    $p_{,X}=0$.     At    the    minimum
$\tau_{,X}(X_*)\equiv  2X_*\,p_{,XX}\,(X_*)$ is  negative,  whereas at
the  origin $\tau_{,X}(0)=1$  is positive.  By continuity  there  is a
point  $X_{**}$   where  $\tau_{,X}$  is   zero,  violating  condition
(\ref{eq:regular}).   We are  forced  to choose  $p\approx -X$  around
$X_\infty=0$.   An analogous  argument implies  that the  function $p$
cannot have  a local maximum, so  we conclude that $p(X)$  should be a
monotonically decreasing function which  is linear at the origin.  The
simplest function which satisfies this criterion is $p=-X$.

\section{Wormhole solutions}\label{sec:solutions}

Our  previous analysis  has  determined that  $p=-X$  is the  simplest
Lagrangian which may allow a  wormhole solution. From now on, we shall
concentrate  on  this form  of  the  Lagrangian.   Notice that  for  a
conventional  massless scalar field  $p=X$.  Hence,  the field  we are
considering is a massless field with a reversed sign kinetic term.  As
long as the effect of the field on gravity can be ignored, the overall
sign  of  the  matter  Lagrangian  is  a  matter  of  convention.   In
particular,  in any  given  gravitational background  the equation  of
motion  of such  a field  is the  same as  the one  of  a conventional
massless scalar, $\nabla^\alpha\nabla_\alpha\varphi=0$.  It is the way
the  field  couples  to  gravity  that  determines  the  sign  of  the
Lagrangian.  If  the scalar field  couples only to gravity,  this sign
can  be  chosen  at  will,  as   long  as  this  does  not  yield  any
inconsistency.

Static,  spherically  symmetric   solutions  of  Einstein's  equations
minimally coupled to a  \emph{conventional} massless scalar field have
been   repeatedly  discovered   in  the   literature   \cite{Fisher}.  
Particularly useful for our purposes are the solutions as presented by
Wyman (see also \cite{AgneseLaCamera}):
\begin{equation}\label{eq:Wyman}
  ds^2=e^{-2m/\tilde{r}}dt^2-e^{2m/\tilde{r}}
  \left(\frac{\eta/\tilde{r}}{\sinh(\eta/\tilde{r})}\right)^4 
  d\tilde{r}^2-e^{2m/\tilde{r}}\left(\frac{\eta/\tilde{r}}
    {\sinh(\eta/\tilde{r})}\right)^2 \tilde{r}^2 d\Omega^2,
\end{equation}
where $\eta^2=m^2+q^2$  and $q$ is defined  by the form  of the scalar
field solution
\begin{equation}\label{eq:field}
  \varphi=\frac{q}{\tilde{r}}.
\end{equation}
These solutions are characterized by two integration constants, a mass
$m$  and  a  ``charge''   $q$.   Because  the  scalar  field  solution
(\ref{eq:field}) is proportional to $q$ and the energy-momentum tensor
(\ref{eq:EMT})  is quadratic  in  $\varphi$, a  real  $q$ describes  a
conventionally coupled  scalar field, while a purely  imaginary $q$ is
equivalent to considering a field with a negative kinetic term. For an
ordinary coupling  ($q^2>0$) the above solutions  describe a spacetime
with  a  naked  singularity   at  $\tilde{r}=0$.   In  fact,  after  a
coordinate transformation, (\ref{eq:Wyman}) can be cast in the form
\begin{equation}\label{eq:LaCamera}
  ds^2=\left(1-\frac{2\eta}{r}\right)^{m/\eta}dt^2-
  \left(1-\frac{2\eta}{r}\right)^{-m/\eta}dr^2-
  \left(1-\frac{2\eta}{r}\right)^{1-m/\eta}r^2 d\Omega^2, 
\end{equation}
where one sees that the would-be Schwarzschild horizon has shrunk to a
point  at   $r=2\eta$.   If  the  charge   vanishes,  $q=0$,  equation
(\ref{eq:LaCamera}) obviously reduces to the Schwarzschild metric.

If $\eta^2=m^2+q^2$ itself is negative, the metric (\ref{eq:LaCamera})
is not well defined.  Of course,  $\eta^2$ can be negative only if $q$
is purely  imaginary (we  assume the  mass $m$ to  be real.)   This is
physically equivalent  to a real  scalar field whose kinetic  term has
the opposite sign,  which is precisely the coupling  we are interested
in.  In what  follows,  we assume  that  both $\eta^2$  and $q^2$  are
negative.    By   a  change   of   radial   coordinates,  the   metric
(\ref{eq:Wyman}) can be rewritten as
\begin{equation}\label{eq:wormhole-set}
  ds^2=e^{-2m/\tilde{r}}dt^2-e^{2m/\tilde{r}}
  \left[dr^2+(r^2-\eta^2)\,d\Omega^2\right],
\end{equation}
where the old coordinate  $\tilde{r}$ is implicitly expressed in terms
of the new coordinate $r$ by the relation
\begin{equation}
  \sin^2\left(\frac{|\eta|}{\tilde{r}}\right)=
  \frac{-\eta^2}{r^2-\eta^2},
\end{equation}
and the explicit functional dependence $\tilde{r}(r)$ is determined by
the  requirements  of continuity  and  differentiability.  The  metric
(\ref{eq:wormhole-set}) was first discovered in a different coordinate
system  in  \cite{Ellis} (see  also  \cite{Bronnikov}), and  classical
scattering    in   such    a    geometry   has    been   studied    in
\cite{ChetouaniClement}.

\begin{figure}
  \begin{center}
    \includegraphics{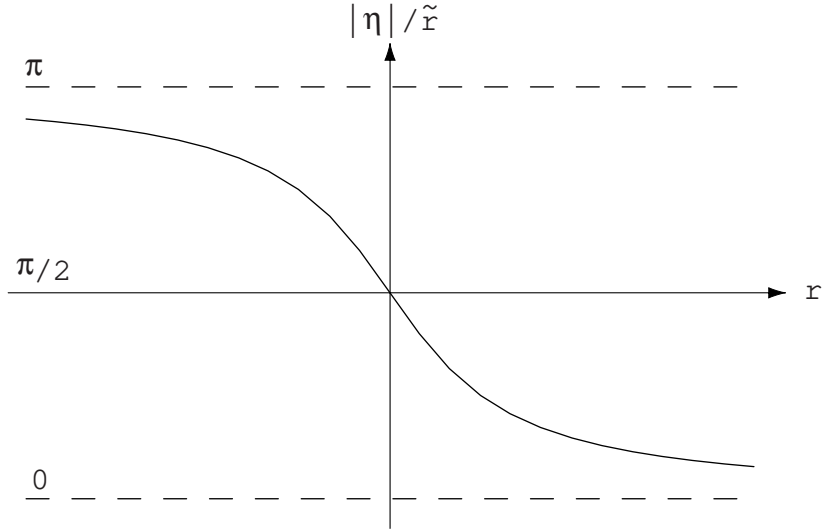}
    \caption{The form of $|\eta|/\tilde{r}$ as a function of
      the coordinate $r$. \label{fig:r-of-r}} 
  \end{center}
\end{figure}

Although we shall not  cast (\ref{eq:wormhole-set}) in proper distance
coordinates,  it  is  quite  clear  that it  describes  a  traversable
wormhole.  Due to  the presence of the $-\eta^2$  term, the coordinate
$r$ may take values in  the range $-\infty\leq r\leq\infty$.  The form
of  $|\eta|/\tilde{r}$  as  a  function  of $r$  is  shown  in  Figure
\ref{fig:r-of-r}.    The  function  $\exp(-2m/\tilde{r})$   is  finite
everywhere and in  particular, the metric (\ref{eq:wormhole-set}) does
not   contain   any   horizon.     In   the   limit   $r\to   \infty$,
$\exp(-2m/\tilde{r})=  1-2m/r+\mathcal{O}(r^{-2})$, and  in  the limit
$r\to    -\infty$,   $\exp(-2m/\tilde{r})\to    \exp(-2\pi   m/|\eta|)
(1-2m/r+\mathcal{O}(r^{-2}))$.    Hence,  the   spatial   sections  of
(\ref{eq:wormhole-set})  contain two  asymptotically flat  regions and
the masses  of the wormhole as  observed in those regions  are $m$ and
$-m\exp(\pi m/|\eta|)$ respectively.  Notice  that the two masses have
opposite  sign.  In  addition, because  the redshift  function  is not
symmetric under reflections $r\to  -r$, clocks tick at different rates
in both  asymptotic regions  and a photon  emitted at  $r=\infty$ will
appear to be  blue shifted at $r=-\infty$.  The  $r$ dependence of the
scalar field also has the form of Figure \ref{fig:r-of-r}.  The scalar
field solutions is  proportional to $q$ and that is  the reason why it
can be  interpreted as  a scalar charge  (this identification  will be
made more precise below.)

The  zero mass  wormhole  solutions  have a  remarkably  simple form.  
Setting $m=0$ in (\ref{eq:wormhole-set}) and using (\ref{eq:field})
to compute $p=-\frac{1}{2}(\nabla \varphi)^2$ one gets
\begin{subequations}\label{eq:zero-mass}
  \begin{eqnarray}
  ds^2&=&dt^2-dl^2-(l^2-q^2)\,d\Omega^2, \label{eq:zero-mass-metric}\\
  p&=&\tau=\frac{1}{2}\frac{-q^2}{(l^2-q^2)^2}.
  \end{eqnarray}
\end{subequations}
Recall that  $q^2$ is negative for the  solutions under consideration,
that is,  the pressure and  radial tension are positive,  in agreement
with  the fact  that $p=-X>0$.   The  reader can  easily verify  that,
indeed,   (\ref{eq:zero-mass})  satisfy  Einstein's   field  equations
(\ref{eq:einstein}).   The  static metric  (\ref{eq:zero-mass-metric})
has  the  wormhole   form  (\ref{eq:metric}).   Its  spatial  sections
$t=const$ contain  two asymptotically flat regions at  $l=\pm \infty$. 
The minimal area of a $l=const$  surface occurs at $l=0$, and is given
by $-4\pi q^2$.  The redshift  function is constant everywhere, and in
particular, there are no tidal  forces which could prevent the passage
of a  person through the wormhole  if $|q|$ is  sufficiently big.  The
metric  in (\ref{eq:zero-mass})  is in  fact the  simplest traversable
wormhole geometry one may think of.

Suppose  that observers  living in  (\ref{eq:zero-mass}) were  able to
measure the  value of  the scalar field.   They would realize  that it
obeys  the  equation  $\Delta\varphi\propto  (r^2  \varphi')'=0$.   By
computing the flux of the  field gradient through surfaces of constant
$l$, they would conclude that this flux is independent of the value of
$l$ and is given by $4\pi|q|$.  If these observers did not have enough
resolving power  to explore  lengths of size  $|q|$, they  would hence
arrive at the conclusion that there is a scalar source of charge $|q|$
sitting within $l<q$.  However, there is no scalar source present.  In
reality  there  is a  sourceless  incoming  flux  which originates  at
$l=-\infty$, passes through the  wormholes mouth at $l=0$, and emerges
again for $l>0$ as an  outgoing flux reaching $l=\infty$.  This is one
example of Wheeler's ``charge without charge'' \cite{Wheeler}.

\section{Wormhole stability}\label{sec:stability}

One of the advantages of  a microscopic description of the matter that
threads a  wormhole is  that such a  description allows us  to address
issues which  would have not been treatable  otherwise.  The stability
of  wormholes is such  an example.   In the  following, we  analyse in
first  order  perturbation  theory   the  stability  of  the  wormhole
solutions  we have  previously  discussed.  Because  of their  extreme
simplicity,    we   concentrate   on    the   zero    mass   solutions
(\ref{eq:zero-mass}),  although we  shall later  argue  that solutions
with  sufficiently small $m$  are also  stable.  Our  analysis closely
resembles an analogous  investigation in \cite{ReggeWheeler} (see also
\cite{Vishveshwara} and \cite{Zerilli}), although  we will not work in
Schwarzschild coordinates but in the proper distance coordinate system
(\ref{eq:metric}),   where  all   metric   coefficients  are   regular
functions.

The  strategy  is  the  following.   We  introduce  metric  and  field
perturbations,  $g_{\alpha\beta}\to  g_{\alpha\beta}+h_{\alpha\beta}$,
$\varphi\to  \varphi+\delta\varphi$  around  the  background  solution
(\ref{eq:zero-mass}).   Linearizing Einstein's  equations  one gets  a
closed  system of equations  for the  metric and  field perturbations,
$\delta  G_{\alpha\beta}=\delta T_{\alpha\beta}$.  If  these equations
admit spatially  well behaved  solutions, that is,  perturbations that
decay sufficiently fast at  spatial infinity, which grow exponentially
in time, our background solutions are unstable.  The symmetries of the
background  solution  allow  us   to  simplify  the  analysis  of  its
perturbations.   Due  to  the  staticity  of the  background,  we  can
decompose the  perturbations into Fourier modes  proportional to $e^{i
  \omega t}$ which  do not couple to each other in  the linear regime. 
Similarly,  because the  background  is spherically  symmetric, it  is
convenient  to decompose  the perturbations  into  spherical harmonics
$Y_{Lm}(\theta, \phi)$  and its  derivatives, and again,  the symmetry
under  rotations makes  possible to  consider only  perturbations with
azimuthal eigenvalue  $m=0$.  Finally,  the inversion symmetry  of the
background  ($l\to  l$,  $\theta\to \pi-\theta$,  $\phi\to  \pi+\phi)$
allows  us to  consider  perturbations of  definite  parity $P$  under
inversion.   Perturbations   of  ``electric''  (or   even)  type  have
$P=(-1)^{L}$  and perturbations  of  ``magnetic'' (or  odd) type  have
parity    $P=(-1)^{L+1}$.    In   linearized    perturbation   theory,
perturbations with different ``quantum numbers'' $\omega$, $L$ and $P$
decouple from each other.

In  Regge-Wheeler  gauge,   the  electric  type  metric  perturbations
(denoted by $\mathrm{(e)}$) are
\begin{equation}\label{eq:metric-even}
h^\mathrm{(e)}_{\alpha\beta}=\left(
\begin{array}{cccc}
H_0(l) & H_1(l) & 0 & 0 \\
H_1(l) & H_2(l) & 0 & 0 \\
0 & 0 & r^2 K(l) & 0 \\
0 & 0 & 0 & r^2  K(l)  \sin^2 \theta
\end{array}
\right) P_L(\cos \theta) e^{i\omega t}.
\end{equation}
The  indices $\alpha$  and $\beta$  run  over the  coordinates $t,  l,
\theta$  and  $\phi$.  The  Legendre  polynomial  $P_L$  is the  $m=0$
spherical harmonic  $Y_{L\,m}$. As previously mentioned,  the time and
angular dependence of the  perturbations has been explicitly separated
and hence,  only their  dependence on the  $l$ variable remains  to be
determined.   Similarly, the  magnetic perturbations  (denoted  by the
label $\mathrm{(m)}$) are
\begin{equation}\label{eq:metric-odd}
h^\mathrm{(m)}_{\alpha\beta}=\left(
\begin{array}{cccc}
0 & 0 & 0 & h_0(l) \\
0 & 0 & 0 & h_1(l) \\
0 & 0 & 0 & 0 \\
h_0(l) & h_1(l) & 0 &  0
\end{array}
\right) \sin \theta \frac{d}{d\theta} 
P_L(\cos \theta) e^{i\omega t},
\end{equation}
so  that there  are  4  electric and  2  magnetic metric  perturbation
functions in total.  The remaining  4 functions needed to complete the
most general metric perturbation are zero in Regge-Wheeler gauge.  The
scalar field perturbations are
\begin{equation}
  \delta\varphi^\mathrm{(e)}=\delta\varphi^\mathrm{(e)}(l)P_L(\cos \theta)
  e^{i\omega t}\quad 
\rm{and} \quad \delta\varphi^\mathrm{(m)}=0,
\end{equation}
and in  particular, because  a scalar field  is a scalar  under parity
transformations, there  are no magnetic type  field perturbations.  It
turns out that  both for electric and magnetic  perturbations, one can
derive  an equation  for a  single perturbation  variable  $Q^{\rm (e,
  m)}$.   The equation  has  the form  of  a Schr\"odinger  eigenvalue
problem
\begin{equation}\label{eq:Q}
  -\frac{d^2}{dl^2}Q^{{\rm (e, m)}}+V_L(l) Q^{{\rm (e, m)}}=
  \omega^2 Q^{{\rm (e, m)}},
\end{equation}
where the potential $V_L$ also  depends on the angular momentum $L$ of
the   perturbations.   Solutions  of   this  equation   with  negative
$\omega^2$  correspond to  exponentially growing  and decaying  modes. 
Hence,  our task  consists  in verifying  whether (\ref{eq:Q})  admits
spatially well behaved solutions  of negative $\omega^2$ . Because the
potential  goes  to 0  as  $l\to\pm  \infty$,  this is  equivalent  to
verifying  whether   $V_L$  admits  any  bound  state,   that  is,  an
eigenfunction of negative energy.

\paragraph{Electric perturbations}
For these perturbations, the relevant linearized equations read
(dropping the (e)-label)
\begin{subequations}\label{eq:electric-linearized}
  \begin{align} 
    \label{eq:even-tt}
    &&[L(L+1)+2] H_2+2 H_2' r r'+(L+2)(L-1)K-
    6K' r r'-2 K'' r^2=r^2\varphi' \delta\varphi'\\
    \label{eq:even-tl}
    &&L(L+1)H_1-2i \omega K r r'+2i\omega H_2 r r'
      -2i\omega K' r^2=i\omega r^2 \varphi'\delta\varphi\\ 
    \label{eq:even-ll}
    &&2H_2-L(L+1)H_0+(L+2)(L-1)K
    +(2H_0'-2K'-4i\omega H_1) r r'-\hspace{2cm}\nonumber \\
    &&{}-2\omega^2 K r^2=-r^2\varphi'\delta\varphi'\\
    \label{eq:even-lx}
    &&\frac{d P_L}{d\cos \theta}\left[H_0 r'-
      H_2 r'-H_0'r+K' r +i \omega H_1 r\right]=
    -\frac{d P_L}{d\cos \theta} r \varphi'\delta\varphi\\
    \label{eq:even-xx}
    &&\frac{d^2 P_L}{d(\cos \theta)^2}\cdot(H_0-H_2)=0\\  
    \label{eq:even-field}
    &&-\omega^2\delta\varphi-\delta\varphi''-
    2\frac{r'}{r}\delta\varphi'+\frac{L(L+1)}{r^2}\delta\varphi
    -\frac{1}{2}(H_0'+H_2'-2K'-2i\omega H_1)\,\varphi'=0,
  \end{align}
\end{subequations}
where  the background  equations (\ref{eq:einstein})  have been  used. 
The cases  $L=0$, $L=1$ and  $L\geq2$ require separate  treatment.  If
$L\geq  2$   equation  (\ref{eq:even-xx})  yields   $H_0=H_2$.   Then,
equation  (\ref{eq:even-lx})  can  be   used  to  express  the  metric
perturbations    of    the    linearized   scalar    field    equation
(\ref{eq:even-field})   in    terms   of   the    field   perturbation
$\delta\varphi$.  The  latter equation reduces  to (\ref{eq:Q}), where
the potential is given by
\begin{equation}\label{eq:even-potential}
  V_L(l)=\frac{L(L+1)}{r^2}-3\frac{r''}{r}=
  \frac{L(L+1)\,l^2-q^2(L^2+L-3)}{(l^2-q^2)^2},
\end{equation}
and $Q^{\rm  (e)}\equiv r \delta\varphi$. For $L\geq  2$ the potential
is  positive everywhere.   Because  positive potentials  do not  admit
bound  states ($E<0$),  it is  clear  that all  electric modes  with
$L\geq 2$ are stable.

\begin{figure}
  \begin{center}
    \includegraphics{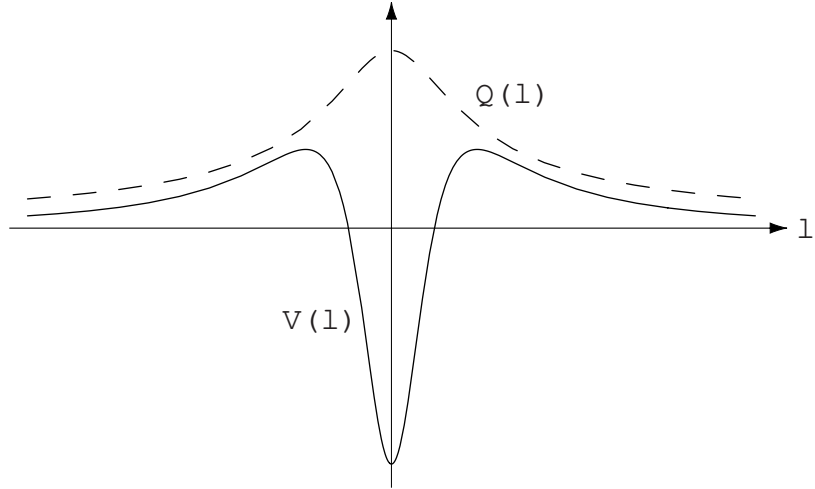}
    \caption{The $L=1$ potential for even perturbations
      (continuous) and its ground state (dashed).\label{fig:L=1}}
  \end{center}
\end{figure}

Although   for  $L=1$   the  linearized   equation  (\ref{eq:even-xx})
identically vanishes,  it is  still possible to  choose a  gauge where
$H_0=H_2$, since  $L=1$ perturbations  have fewer free  functions than
$L\geq  2$  perturbations.   The  potential  is thus  still  given  by
(\ref{eq:even-potential}), but  in the latter  case it has  a negative
minimum at  $l=0$.  The $L=1$ Schr\"odinger  equation (\ref{eq:Q}) can
be  solved  exactly  for   $\omega^2=0$.   One  solution  is  Meijer's
G-function
$G^{11}_{22} \left(1-l^2/q^2\,\big|{\tiny\begin{array}{cc} 1/2 &  2 \\
      3/2 & -1/2  \end{array}}\!\right)$ \cite{GraRyz}.  This solution
is shown in Figure \ref{fig:L=1}.   It is even, decays at infinity and
it has no nodes.  Hence it  is the ground state of the potential $V_1$
\cite{MorseFeshbach}.    In   particular   there  are   no   spatially
well-behaved solutions  at infinity  with $\omega^2<0$, and  the $L=1$
mode is stable.

The most general metric perturbation for $L=0$ (spherically symmetric)
has fewer free  functions than for $L\geq 1$. In  fact, it is possible
to  choose a gauge  where $K=H_1=0$.   Using the  linearized equations
(\ref{eq:even-tt}),    (\ref{eq:even-tl})    and   (\ref{eq:even-ll}),
equation (\ref{eq:even-field})  can be cast in  the form (\ref{eq:Q}),
where the potential is given by
\begin{equation}
  V_{0}(l)=\frac{r''}{r}+2\frac{r''}{r\,r'{}^2}=
  \frac{2\,q^4-3\,q^2\, l^2}{l^2\,(l^2-q^2)^2}.
\end{equation}
The  latter  potential is  everywhere  positive,  and hence,  equation
(\ref{eq:Q})  does  not admit  any  $\omega^2<0$ solutions.   Electric
spherically symmetric perturbations are also stable.

\paragraph{Magnetic perturbations} 

The only  non-trivial linearized Einstein field  equations it suffices
to consider are the $x\phi$ and $l\phi $ respectively,
\begin{subequations}\label{eq:magnetic-linearized}
  \begin{gather}
    \label{eq:mag-xphi}
    (h_1'-i\omega h_0)\frac{d^2 P_L}{d(\cos \theta)^2}=0 \\
    \label{eq:mag-lphi}
    -\omega^2h_1-i\omega h_0'+2i\omega\frac{r'}{r}h_0+
    2\frac{r''}{r}h_1+\frac{(L+2)(L-1)}{r^2}h_1=2\, \varphi'{}^2\, h_1.
  \end{gather}
\end{subequations}
There  are no magnetic  perturbations with  $L=0$. For  $L=1$ equation
(\ref{eq:mag-xphi}) does  not contain significant  information and the
remaining equation,  (\ref{eq:mag-lphi}), provides a  relation between
$h_1$ and $h_0$ that allows  us to gauge these perturbations away. For
$L\geq2$    equation   (\ref{eq:mag-xphi})    yields    the   relation
$h_1'=i\omega  h_0$, which  when substituted  into (\ref{eq:mag-lphi})
allows us to put it in the form (\ref{eq:Q}), where $Q^{\rm (m)}\equiv
h_1/r$ and  the potential is  the same as for  electric perturbations,
(\ref{eq:even-potential}).   Therefore, $L\geq  2$ magnetic  modes are
also stable.

Up to  now, our discussion  has focused on  the stability of  the zero
mass  solutions.   Because  of  that,  we have  only  considered  time
dependent perturbations,  i.e on non-zero values of  $\omega$.  If one
studies  time-independent  $L=0$  perturbations,  one has  to  recover
non-zero mass solutions (\ref{eq:wormhole-set}) for sufficiently small
values  of  $m$ as  time-independent  perturbations  of the  zero-mass
solutions.  Due to  linearity, the time-dependent perturbations around
these   non-zero  mass   solutions  satisfy,   of  course,   the  same
perturbation      equations     (\ref{eq:electric-linearized})     and
(\ref{eq:magnetic-linearized}).  Since these equations do not have any
spatially well-behaved growing solution, we conclude that at least for
sufficiently small values of $m$ the solutions (\ref{eq:wormhole-set})
are also stable. This completes our wormhole stability analysis.

Up to  some extent,  the stability of  the wormhole solutions  we have
considered  was  to  be  expected  in  the light  of  the  results  of
\cite{GaMu}.  In that work, it  is shown that the speed of propagation
of linear perturbations in  a cosmological background, $c_s$, is given
by
\begin{equation}
c_s^2=\frac{p_{,X}}{\rho_{,X}}.
\end{equation}
If the squared speed of  happened to be negative, one would anticipate
instabilities of  the wormhole associated with  the exponential growth
of short-wavelength modes.  Because $c_s^2=1>0$ in our model, there is
no reason to expect  those instabilities.  However, this argument only
addresses  the issue  partially, since  static,  spherically symmetric
solutions of Einstein's equations  coupled to a \emph{canonical} field
($c_s^2=1$) \emph{are}  unstable \cite{JetzerScialom}. (In  the latter
case the instability appears in the $L=0$ mode.)

\section{Conclusions}

Non-canonical scalar field Lagrangians allow the existence of wormhole
solutions in general relativity. The conditions of asymptotic flatness
and  the  existence  of   a  wormhole  throat  essentially  force  the
non-canonical  scalar field  to  be  a massless  scalar  field with  a
reversed sign Lagrangian.  Zero  mass wormhole solutions to Einstein's
equations coupled to such a field  turn out to be extremely simple and
are  useful as  toy models  to explore  different aspects  of wormhole
physics.  As  a particular application,  it is possible to  show their
stability against linear  perturbations analytically.  The issue about
the  physical viability  of  such  a scalar  field  is yet  unsettled,
although at the level of  our investigation we have not discovered any
internal  inconsistency.  Rather,  it turns  out that  because  such a
field violates  some of  the standard energy  conditions, cosmological
and spherically symmetric solutions  of Einstein's field equations are
singularity-free.

\begin{acknowledgments}
  It is  a pleasure  to thank Sean  Carroll's group,  Stefan Hollands,
  David  Kutasov  and  Robert  Wald  for  constructive  criticism  and
  discussions  on the  subject, and  Jennifer Chen  for  improving the
  original  manuscript.   The author  has  also  benefitted from  many
  useful conversations over the time with Slava Mukhanov.  Some of the
  calculations in  this paper have been performed  with the GRTensorII
  package for Mathematica.   This work has been supported  by the U.S. 
  DoE grant DE-FG02-90ER40560.
  
  I would like  to thank also G.  Clement and  K. Bronnikov for kindly
  bringing         to          my         attention         references
  \cite{Ellis,Bronnikov,Kodama,ChetouaniClement}    after    a   first
  submission of the preprint.

\end{acknowledgments}

\appendix

\section{Stability of Minkowski space}\label{sec:appendix}
Our  work has  mainly  focused on  a  massless scalar field described
by a Lagrangian $p$ whose sign has been reversed. In other words,
we have considered a Lagrangian
\begin{equation}\label{eq:twosigns}
p=\frac{\kappa}{2}\nabla_\alpha\varphi\,\nabla^\alpha\varphi
\end{equation}
where $\kappa$  is $-1$ instead of  $+1$.  One of  the main objections
against such an  assumption is that for such  a scalar field, Minkowski
space should be unstable. Indeed, on general grounds one expects small
scalar  field  fluctuations to  radiate  positive  energy  in form  of
gravitational waves, making the negative energy density of the initial
field  fluctuations  even more  negative  and  leading  to a  complete
instability of  the vacuum.  Let us verify  whether one  observes this
kind of  behavior in perturbation  theory.  In order to  establish the
difference  with respect  to an  ordinary scalar  field, we  will keep
$\kappa$  as a  free parameter.   Consider perturbations  of Minkowski
space to second order,
\begin{displaymath}
  g_{\alpha\beta}=\eta_{\alpha\beta}+h^{(1)}_{\alpha\beta}
+h^{(2)}_{\alpha\beta}+\cdots,
\end{displaymath}
where  the notation  should be  obvious, and  consider  equally second
order perturbations of of a constant massless scalar field,
\begin{displaymath}
\varphi=\varphi^{(0)}+\varphi^{(1)}+\varphi^{(2)}+\cdots,
\end{displaymath}
where $\partial_\alpha\varphi^{(0)}=0$. To first order in perturbation
theory, the Einstein and Klein-Gordon equations are
\begin{eqnarray}
  \eta^{\mu\nu}\partial_\mu\partial_\nu
  \left(h^{(1)}_{\alpha\beta}-\frac{1}{2} 
  \eta_{\alpha\beta}h^{(1)}\right)&=& 0, \\
  \eta^{\mu\nu}\partial_\mu\partial_\nu\varphi^{(1)}&=& 0.\label{eq:1-KG}
\end{eqnarray}
Upon gauge fixing, the  first of the equations describes gravitational
waves  propagating  in  Minkowski  spacetime,  while  the  second  one
describes  scalar field  waves  propagating in  the  same background.  
Notice that  these equations  are the same  regardless of  $\kappa$ in
(\ref{eq:twosigns}).   At this level,  both signs  are equally  valid. 

Consider then the second order Einstein equations,
\begin{equation}\label{eq:backreaction}
\eta^{\mu\nu}\partial_\mu\partial_\nu\left(h^{(2)}_{\alpha\beta}
  -\frac{1}{2}\eta_{\alpha\beta}h^{(2)}\right)
=4 \kappa  \left(\partial_\alpha \varphi^{(1)}
  \partial_\beta \varphi^{(1)}-\frac{1}{2}\eta_{\alpha\beta}
  \partial_\gamma\varphi^{(1)}\partial^\gamma\varphi^{(1)}\right)
-2 G^{(2)}(h^{(1)}).
\end{equation}
The sign  of the  first term on  the right  hand is determined  by the
conventional/unconventional  coupling of  the massless  scalar  field to
gravity and the second term on  the right hand side contains all terms
of  the Einstein  tensor quadratic  in the  first  order perturbations
$h^{(1)}$.  Thus, the right hand  side of the above equation describes
how the first order  scalar field and gravitational waves respectively
back-react on  the background geometry. Whereas  the gravitational wave
back-reaction is the same for  both signs of the scalar field coupling,
the scalar field  wave back-reaction on the metric  does depend to this
order  on the sign  of the  energy momentum  tensor, as  expected. The
Klein-Gordon equation to second order
\begin{displaymath}
\eta^{\mu\nu}\partial_\mu\partial_\nu \varphi^{(2)}=-\partial_\alpha
\left(h^{(1)}{}^{\alpha\beta}\partial_\beta\varphi^{(1)}
-\frac{1}{2}h^{(1)} \partial^\alpha\varphi^{(1)}\right)
\end{displaymath}
describes how first order scalar and gravitational waves generate
second order field perturbations. For given perturbations, it
is the same for both signs of the scalar field coupling.

Suppose  that we  solve  equation (\ref{eq:backreaction})  in a  given
background of  first order  scalar waves (assume  for the sake  of the
argument  that there  are no  gravitational waves.)   The  solution is
given by the convolution of the (known) source terms in the right hand
side with  the appropriate retarded Green's  function.  In particular,
the solutions for  both values of $\kappa$ differ only  by a sign.  If
the solution is  well behaved for $\kappa=1$, it  is also well behaved
for $\kappa=-1$.  Thus, at second  order there is not yet any evidence
that $\kappa=+1$ is preferable to $\kappa=-1$.  It seems  that if we
want to  single out $\kappa=1$ as  the preferred choice  of the scalar
field coupling, we have to go to higher orders.

Due  to  the structure  of  the  perturbation  equations, even  if  we
proceeded  to higher  orders it  would  be difficult  to discover  why
$\kappa=+1$  is  essentially different  from  $\kappa=-1$  (as in  the
``analogous''  system  of  two  coupled  scalar  fields,  one  with  a
conventional, the other with an unconventional kinetic term.)  At this
point let us  try a non-perturbative analysis and  point out that even
the conventional coupling $\kappa=1$ leads to instabilities.  Consider
for that  purpose an homogeneous  spatially flat universe filled  by a
canonical scalar field,
\begin{displaymath}
ds^2=dt^2-a(t)^2\left[dr^2+r^2d\Omega^2\right].
\end{displaymath}
It is known that the solution of Einstein's and Klein-Gordon equations
in such a spacetime is given by
\begin{eqnarray*}
a(t) &\propto& |t|^{1/3} \\
\left(\frac{d\varphi}{dt}\right)^2 &\propto& \frac{1}{t^2}.
\end{eqnarray*}
This  solution  has  two  branches,  related to  each  other  by  time
reversal. For positive times, the universe starts expanding at the big
bang  singularity  at  $t=0_+$   and  approaches  Minkowski  space  at
$t=\infty$ (the  Hubble parameter $d\log a/dt$  approaches zero).  For
negative times the universe starts from Minkowski space at $t=-\infty$
and  contracts into a  ``pre-big bang''  singularity at  $t=0_-$. This
latter  branch  implies that  a  canonical  massless  scalar field  in
Minkowski  space is  ``unstable'' upon  contraction.  As  a  matter of
fact, this  instability is  one of the  ideas behind the  pre-big bang
scenario  and  reflects  nothing  else other  than  the  gravitational
instability upon collapse of the scalar and gravitational waves around
Minkowski   spacetime   \cite{DamourVeneziano}   we  have   previously
encountered.  Notice  that an  analogous argument applies  for $p=-X$,
though  there is a  crucial difference  too. In  the latter  case, the
scalar field has  a negative energy density, and  the spatial sections
of  the universe have  to be  negatively curved.   Expanding solutions
hence asymptotically approach the  ``Milne universe'', which is just a
portion of  Minkowski space. On  the other hand  contracting solutions
(the time reversed expanding solutions) originate from Minkowski space
and instead of running into a singularity, bounce at a finite value of
$a$  and  expand again  into  Minkowski  spacetime  (this is  possible
because    our   field    violates   the    null    energy   condition
\cite{VisserMolina}).

\end{document}